# USER EXPERIENCE WITH EDUCATIONAL TECHNOLOGY IN AFRICAN SLUMS

G. Stefansson, A. H. Jonsdottir

*University of Iceland, Science Institute (ICELAND)*

## Abstract

This paper describes a project developed in co-operation with two dozen community libraries and schools in various slums and low-income regions in Kenya. The project was started in response to COVID-19, to allow students to solve computerised math drills while schools were closed. The number of students involved reached two thousand during the first 24 months of operation. The program uses a study environment, tutor-web, and access to this is provided by donating tablet computers to participating community libraries. Students are rewarded using tokens, SmileyCoins or SMLY, as they progress through the system and the libraries are free to sell for SMLY small food items, sanitary pads and even the tablets themselves. The rewards are designed to put an emphasis on secondary school mathematics, so as to prepare the students for applications into STEM subjects at university. Completion of the corresponding collection of drills gives SmileyCoin awards sufficient to purchase a tablet.

In order to investigate the students' experience, engagement and the effects of rewards, a survey was built into the tutor-web system. Conclusions based on these first two years indicate that most students find the system easy to use, learn quite a bit from using it, but their underlying reasons for participating are mixed. The majority of the students express a desire to go to university.

## 1 INTRODUCTION

The tutor-web system [1] has been used for over a decade at multiple schools and universities in Iceland and Kenya, primarily for the purpose of teaching mathematics and statistics. The mobile-web is freely accessible to all learners and can be accessed via the website http://tutor-web.net. Upon utilizing the tutor-web, learners are presented with exercises that are tailored to their current level of understanding. Following the submission of their answers, detailed feedback is provided. The system offers a vast array of exercises in mathematics and statistics, covering both high school and university level material. It is worth noting that users are not required to have an internet connection when answering exercises, only when downloading the exercise banks to their devices. An illustration of an exercise and the feedback provided to students upon submission (yellow) can be seen in Figure 1. The tutor-web system offers several noteworthy features that will not be discussed in depth in this paper. These include the capability of incorporating real-world data into exercises, either through user uploads or through integration with open data sources. Additionally, students are incentivized to engage in the system through the opportunity to earn a cryptocurrency, known as the SmileyCoin. Further information regarding these features can be found in [2] and [3]. A comprehensive description of the tutor-web system can be found in [4].

In Kenya, the system is integrated as part of an initiative to improve mathematics education through the utilization of educational technology. The program is organized by the SmileyCharity, a non-profit organization (registered in Iceland as Styrktarfélagið Broskallar) in partnership with the African Maths Initiative and various individuals. Before 2020 it was implemented by providing servers to schools and tablets to students, with 5 servers donated and some 500 tablets in 2014-2020.

The primary objective of the tutor-web project in Kenya is to provide assistance to secondary school students in successfully gaining admission to universities. The most significant challenge in this regard is passing the national examination, known as the KCSE. The tutor-web system includes a general module for secondary school mathematics, but in 2020, a new module was added specifically targeting the KCSE. This was achieved by sourcing questions from an example exam, developing a computer program for each question and using it to generate drill sets of 100 items per question. The entire collection of drill sets was then integrated as a module in the tutor-web, taking into consideration issues related to rote learning [5].

---

The equation of a curve is $y = -\frac{4}{3}x^3 - \frac{33}{2}x^2 - 35x - 9$

Determine the stationary points of the curve and the nature of those stationary points.

✗ a. ◉ The stationary points are $(-5/4; 1111/96)$ and $(-7; -691/6)$. The first is a minimum and the second is a maximum.

✓ b. ○ The stationary points are $(-5/4; 1111/96)$ and $(-7; -691/6)$. The first is a maximum and the second is a minimum.

c. ○ The stationary points are $(-53/20; 1111/96)$ and $(-7; -691/6)$. Both are minimum points.

d. ○ The stationary points are $(-5/4; 1111/96)$ and $(-20/3; -691/6)$. Both are maximum points.

Stationary point of a curve is a point where the curve stops increasing or decreasing. That is a point where the tangent to the curve is a horizontal line (has slope equal to zero). The slope of the tangent is the derivative of the curve. So to find where the stationary points are we want to solve $\frac{dy}{dx} = 0$. Lets start by calculating $\frac{dy}{dx}$

$$\frac{dy}{dx} = -4x^2 - 33x - 35 = (4x+5)(-x-7)$$

Now lets solve $\frac{dy}{dx} = 0$

$$(4x+5)(-x-7) = 0$$

We can clearly see that the solutions are $x = -5/4$ and $x = -7$. Now we know the x-coordinates of the stationary points. To find the y-coordinates we simply plug the x-coordinates into the equation of the curve and solve for y. We get that $y = 1111/96$ and $y = -691/6$ respectively. So the stationary points are

$$(-5/4; 1111/96) \text{ and } (-7; -691/6)$$

Now the only thing that is left is to determine the nature of the stationary points. That is determine weather they are minimum points, maximum points or saddle points.

We can use a second derivative test to determine whether a stationary point is a local maximum or minimum. A stationary point is classified based on whether the second derivative is positive, negative, or zero. If the second derivative at a stationary point is greater than zero the stationary point is a local minimum, if it's less than zero the stationary point is a local maximum and if it's equal to zero than the test is inconclusive. Lets calculate the second derivative of the curve.

$$\frac{d^2y}{dx^2} = -8x - 33$$

Now we simply plug the x-coordinates of the stationary points in to the second derivative and observe if it is greater or less then zero. We get that

$$-8 \times (-5/4) - 33 = -23 < 0 \text{ and } -8 \times (-7) - 33 = 23 > 0$$

So the first point is a maximum and the second point is a minimum.

Figure 1. An exercise in the tuto-web. Feedback provided to students upon submission is in yellow.

This paper describes various results from an anonymous user survey, routinely answered by students as a part of answering a collection of drillsets for the Kenyan Certificate of Secondary School Education (KCSE).

## 1.1   The Library model

As previously discussed elsewhere [6], significant adjustments to the tutor-web project had to be made during the COVID-19 pandemic as schools in Kenya remained largely closed, preventing the direct donation of tablets to students. Community libraries in Kenya, however, remained open and served as a location for students to continue their studies. In response, the SmileyCharity formed partnerships with multiple organizations in Kenya to distribute tablets to selected libraries, the *SmileyLibraries*. This represents a significant shift in strategy, as the tablets are no longer given directly to students by the charity, but instead are purchased by partner organizations in Kenya and donated to partnering libraries under specific conditions. It's worth noting that there are advantages and disadvantages to both methods, whether giving tablets directly to students or to an organization. In the former case, the recipient and end-user are well defined, while in the latter case, the tablet may have more extensive use as it is borrowed by multiple students. Since the beginning of 2020, the system has been implemented at approximately 30 locations across Africa, primarily in Kenya, with a small number of sites in Ethiopia and Burkina Faso.

These considerations led to the implementation of the *Library model*, where tablets are donated to a library which has full discretion on how they are lent to students. The pre-registered student accounts are set up in the tutor-web and assigned to students by the librarian. Each student can then practice for as long as they desire, including towards mastering the complete KCSE drill set.

The tutor-web system offers rewards for students in the form of grades, similar to other drilling systems, but it also provides an additional incentive by rewarding students with an electronic token called SmileyCoin, as mentioned above. SmileyCoin is a cryptocurrency that can be redeemed by the student and used outside of the tutor-web system. The SmileyCoin Fund is an organization that holds a large number of SmileyCoins, which accepts grant applications and has provided funding for the tutor-web with 1-1.5 billion SmileyCoins annually in recent years. Many of the schools that participated in the program prior to 2020 did not have access to the internet, making the blockchain-based SmileyCoin reward system inaccessible. However, this changed in 2020 as the SmileyLibraries all have local WiFi and internet access, allowing students to earn SmileyCoin rewards as they progress and obtain high grades in the tutor-web. The rewards are not based on prior knowledge, but rather on consistent study and solving of exercises. For a reward system to be effective, the rewards need to make the effort required worthwhile. In earlier SmileyCoin experiments, small online markets were set up where students could purchase items such as discount coupons or movie tickets.

The library model presents a new opportunity, allowing students who have earned sufficient SmileyCoins to purchase the tablet device used in the program. Each student can practice for as long as they desire, towards mastering the complete KCSE drill set, which earns a predefined reward of 1 million SmileyCoins. On the back of each tablet, there is a QR code that corresponds to a payment address. If a student scans this code, they have the option to pay for the tablet using 1 million SmileyCoins. If a library reaches the point where students are purchasing tablets, the SmileyCharity will donate more tablets to the library. At a later stage, the libraries were allowed to set up *Library stores* where students could purchase various small items in exchange for SmileyCoins. The corresponding costs are covered by the SmileyCharity which simply reimburses the librarians for costs corresponding to SmileyCoin deposits to the library address on the blockchain.

The tutor-web system has two types of students: self-registered students and students who are registered as part of a real-world classroom or library. The reward schemes are set differently for these two groups to prevent any misuse of the cryptocurrency.

In this scenario, it is important to exercise caution when determining rewards, as the SmileyCoin Fund has limited resources of SmileyCoins and the SmileyCharity has limited financial resources. The rewards can be adjusted at multiple levels, providing a wide range of options for assigning rewards for performance. Before moving to the Library model, several challenges had to be addressed, including changes to the contract with the funding agency (Ministry of Foreign Affairs in Iceland), establishing agreements with libraries, and coordinating with contact persons and NGOs in Kenya. Videos were created to explain the new approach (e.g. https://bit.ly/TheLibraryModel) and the reward settings in the tutor-web were modified.

Data on student answers to drill items are collected in the tutor-web database, which is typically used by instructors to monitor their students. However, in this setting, there are no instructors, and the data is only available for general, anonymous data analysis. Monitoring scripts were also set up to provide anonymous data on the performance of the libraries, promoting friendly competition among them. The output of these scripts is displayed on a web page at https://libraries.tutor-web.net/.

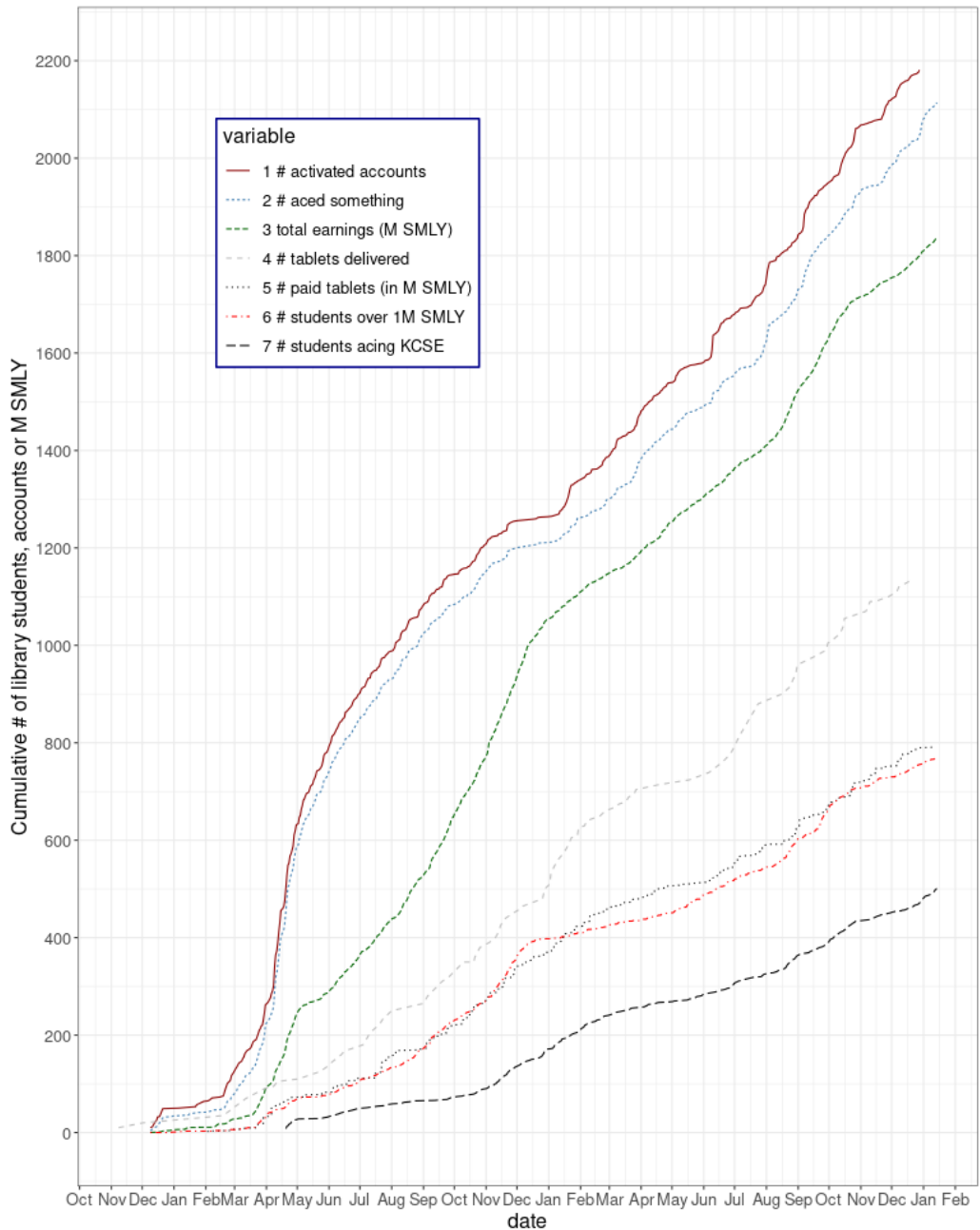

*Figure 2. Cumulative tutor-web use from 2020 into January 2023, earnings and tablet purchases.*

The response to the program is summarized in Figure 2. During the recruitment of new libraries, accounts were distributed widely, usually in batches of 100. Adoption was slow at first, as depicted in the figure, but then saw a significant increase. The figure displays several time series, with the highest curve (1) representing the cumulative number of students who have utilized the tutor-web system and responded to at least one question. The curve (2) which is slightly lower, represents the cumulative number of students who have successfully completed at least one drillset. The green curve (3) illustrates the cumulative amount of SmileyCoin earned by library students in total. The last three curves in the figure relate to tablet accounting and display the count of tablets distributed to libraries, tablets acquired by students, and the number of students who have earned a minimum of one million SmileyCoins and thus able to purchase a tablet.

## 1.2 The data

The KCSE drillset collection consists of over 50 drillsets. The last of these drillsets is a survey with 10 questions. Answers are grouped according to the students's usernames, but these are allocated to students by local librarians and their identities are not made available or recorded outside the libraries, thus remaining anonymous to tutor-web administrators and the survey designers. Further, although the survey is included in the form of multiple-choice drills, students are informed that all responses are marked "correct" and they will not be penalised for checking the option "*I would rather not answer*". By checking this box, they can avoid revealing any of their personal details in order to maintain a grade or receive the SmileyCoin tokens. However, they do need to actually check some answerbox in order to receive the full SmileyCoin award for the full drillset collection. In the tables shown in the Results chapter, the "*I would rather not answer*" and the missing answers are combined in the category "*Missing*".

The survey was placed at the very end of the KCSE collection of drillsets, as the last drillset, in the hope that the students would complete this after completing the true math drills.

These data are obviously not a random sample of all Kenyan students: The tutor-web programme is made available through the selected libraries and the KCSE drillsets will be used only by the students who choose to do so, for whatever reason.

## 2 RESULTS

### 2.1 Composition of the participating learners

A total of 805 students had answered the survey by the end of 2022. Of these, 454 students had aced the KCSE drillset, meaning that they finished each and every one of them with the grade 9.75 or higher.

Achieving gender balance has been difficult in this program with 196 students reporting themselves as female, 406 as male, 39 as other gender, 80 would rather not say and 84 did not answer. The gender proportions for acing the KCSE can be seen in Table 1. The proportions are almost equal in males and females but somewhat lower in learners of other genders. The difference is however not significant (p = 0.218).

*Table 1: Gender proportions for acing the KCSE drillsets.*

|  | *Did not ace the KCSE* | *Did ace the KCSE* |
|---:|:---:|:---:|
| I am a female | 0.403 | 0.597 |
| I am a male | 0.411 | 0.589 |
| Other | 0.538 | 0.462 |
| Missing | 0.338 | 0.338 |

The KCSE material is intended for students in secondary school, but these drills are clearly used by a wider group as can be seen in Table 2.

*Table 2: Level of education of the participation learners.*

|  | **Number of learners** |
|---:|:---:|
| I have not completed primary school | 54 |
| I have completed primary school | 259 |
| I have completed secondary school | 133 |
| I have completed a vocational degree | 51 |
| I have completed a university degree | 60 |
| Missing | 248 |

The 259 students who have completed school and not declared further schooling form the dominant group, but they are not the majority of students who ace these drillsets. There may be many reasons for this, ranging from primary school students wanting to extend their studies to older students who simply need remedial mathematics.

The age distribution of the learners can be seen in Table 3. The majority of the students are between 15 and 18 years old, as expected.

*Table 3: The age distribution of the participating learners.*

|  | **Number of learners** |
|---|---|
| Younger than 12 | 13 |
| 12 - 14 | 100 |
| 15 - 18 | 373 |
| Older than 18 | 103 |
| Missing | 216 |

It is of interest to investigate the students' attitude toward mathematics. The results can be seen in Table 4. Majority of the students like mathematics, some find it "easy" while others find it "sometimes difficult".

These data are provided in a voluntary fashion by those having accounts in the tutor-web system. It follows that some of the responses may not be fully reliable. However, some **sanity checks** can be performed on these data along with **data reduction** to select only those students who are the target group of the educational programme. By selecting only those students who checked "I have completed primary school" **and** "I am currently in secondary school" **and** who had completed the entire collection of KCSE-related drillsets, a **target group** of 126 students is obtained. Table 4 show the same analysis for the target group as well as the full group. Although the columns may be considered quantitatively similar, the difference in frequency distribution between the two sets is highly significant and the "signal" in the target group is much clearer.

*Table 4: Attitude towards mathematics.*

|  | **Number of learners** | **Target group** |
|---|---|---|
| I really like mathematics and I find it easy | 260 | 81 |
| I like mathematics but it is sometimes difficult | 181 | 29 |
| I study mathematics mostly because I have to | 89 | 4 |
| I study mathematics only because I have to | 82 | 4 |
| I try to avoid mathematics | 34 | 1 |
| Missing | 159 | 7 |

In the last two background questions, the learners were asked if they had used a computer before participating in the tutor-web project and how they can access the tutor-web system. When only considering the learners that answered the questions, around 55% of the learners had used a computer before and 45% had not. Around 48% of the learners either own a computer, tablet or smartphone themselves or can access a device owned by a family member.

*Table 5: Computer use before participating in the tutor-web project.*

|  | **Number of learners** |
|---|---|
| I have used a computer or tablet or smartphone before trying the tutor-web | 324 |
| I had never used any computer before I started working in the tutor-web | 274 |
| Missing | 207 |

*Table 6: Ownership of devices.*

|  | **Number of learners** |
|---|---|
| I personally own a tablet or smartphone which I can use to help me with my studies | 148 |
| My family owns a smartphone or tablet which I can sometimes use for studying | 159 |
| I can sometimes access a smartphone or tablet outside the library | 154 |
| I can only access the tutor-web by borrowing a tablet in the library | 185 |
| Missing | 159 |

## 2.2 Learning and usability of the tutor-web system

It is well known, from multiple studies, that students learn while using the tutor-web, as measured from internal tutor-web data and as well as controlled experiments [7]. It is also known from studies in Iceland that students like to use the tutor-web and their own perception is that they learn from using it.

As seen in the following table, the same is true of students in Kenya. By far the most common response is that students learn "a lot" from using the tutor-web and only 12% of true respondents claim they learn nothing from using it.

*Table 7: Learning in the tutor-web.*

|  | **Number of learners** | **Target group** |
|---|---|---|
| I have learned a lot from using the tutor-web | 364 | 90 |
| I have learned a bit from using the tutor-web | 85 | 14 |
| I have not learned much from using the tutor-web | 88 | 11 |
| I have not learned anything from using the tutor-web | 74 | 2 |
| Missing | 194 | 9 |

As in Table 4, note how the signal in the target group is much clearer in spite of the smaller sample size.

The tutor-web is not a fancy commercial website or "app". It is intended for a certain functionality and efforts are mostly expended on adding drills and research-related features rather than ease-of-use. The following question therefore investigates usability.

*Table 8: Usability.*

|  | Number of learners |
|---|---|
| I find the tutor-web easy to use | 325 |
| I sometimes have trouble using the tutor-web | 207 |
| I have a lot of problems using the tutor-web | 98 |
| Missing | 175 |

The single most common response is that students find the web page easy to use (52% of true responses). However, it is also clear that a large proportion (48%) of true responses indicate that students have some problems in using it. Some 16% claim they have "a lot of problems". Given these numbers it would be very useful to conduct a later survey to analyse the exact reasons for the students' troubles. The problems may be design related, lack of local support or the students may simply have a hard time with mathematics and blame it on the tutor-web system.

## 2.3 Incentives

It is known from informal interviews that many students and their families find the SmileyCoin reward scheme important. One question in the survey attempts to disentangle the true motivation of the students. Note that the use of the library stores was not included as an answer option in this question, as these only became important later.

*Table 9: The learner's incentive for using the tutor-web.*

|  | Number of learners |
|---|---|
| I use the tutor-web only to learn, and I have no plans to use the SmileyCoin for anything | 120 |
| I use the tutor-web mainly to learn, but I also plan to to earn enough SmileyCoin to buy tablets | 361 |
| I only use the tutor-web so that I can earn enough SmileyCoin to buy tablets, but I learn quite a bit at the same time | 112 |
| I only use the tutor-web so that I can earn enough SmileyCoin to buy tablets, I don't really learn anything | 54 |
| Missing | 158 |

The most common answer (56%) here is the use of tutor-web for learning while also wanting to use SmileyCoin for purchasing tablets. However, the answers can be grouped and interpreted in a few different ways. Using only the true respondents (ie omitting "Missing"), we see that the tutor-web has a primary use for learning for 74% of the students, 92% learn from using it and 82% of the students want to use their coins.

It is also possible group the responses as if they use SMLY or not and if they think they learn from using the tutor-web or not as done in Table 10. Note that the structure of the original question assumes that the empty cell in the table is truly empty, ie it is assumed in the question itself that anyone using the tutor-web is using it to learn or to earn SmileyCoin (or both).

*Table 10: Use of SMLY and learning.*

|  | **Learn** | **Do not learn** | **Total** |
|---|---|---|---|
| Use SMLY | 473 | 54 | 527 |
| Do not use SMLY | 120 |  | 120 |
| Total | 593 | 54 | 647 |

Answers to the following survey question indicate that a large majority of survey participants want to go to university, or 79% of those who express an opinion.

*Table 11: The learner's interest in going to university.*

|  | **Number of learners** | **Target group** |
|---|---|---|
| I would like to go to university | 524 | 115 |
| I am not interested in going to university | 138 | 6 |
| Missing | 143 | 5 |

Splitting this table according to whether the students have completed the KCSE drillsets or not indicates a highly significant difference in attitude between those two groups ($p < 0.001$). Overall, most students want to go to university and most have completed the KCSE drillsets. However, considering only those who have no interest in going to university, the majority have not completed the drillsets.

The last column of the table contains the subset of students who are the actual target group, i.e. have completed the drillsets as well as indicated that they are in the appropriate stage of their studies. Notice that here 95% of those in the target group who respond express a desire to go to university.

Naturally, there is a "selection bias" in the last two analysis which consider (subsets of) those students who have completed the drillsets: One would expect it to be more likely that those interested in university will complete the drillsets.

## 2.4 Comparison between gender

For various reasons, one cannot make *a priori* assumptions on how well the different genders are prepared for using the tutor-web system. Their background may or may not be different, their exposure to computers may or may not be different, their perceptions of the usefulness of the system may or may not be different and so forth. As shown in Section 2.1, there is no significant gender difference on whether students have completed the KCSE material before answering the survey.

When looking at a potential gender difference in the learner's incentive, learning, usability, attitudes toward math, whether they want to go to university and prior use of tablets, no significant difference was found. In these analyses, only learners that identify themselves as males or females were included because of few answers from learners of other genders. There was one comparison on the verge of being significant ($p = 0.069$), the interest in attending university, with 88% of female learners expressing interest and 82% of male learners.

## 3 CONCLUSIONS AND DISCUSSION

A method is described to enhance mathematics education in Kenyan slums by providing tablets to community libraries and allowing students to purchase the tablets by using a cryptocurrency earned by studying. The resulting effect on participation and performance is unprecedented: Eleven libraries with 1301 students opted for voluntary participation in 2021 causing the program to run at full financial capacity. In that year, 450 students earned enough SMLY to purchase the tablets, which involves completing a large collection of drills to a level of excellence.

While practising in the tutor-web drilling system, the students demonstrate an increased ability in general mathematics ability, measured by a yardstick which is not a part of the drills themselves.

The surveys include questions on gender. The disparity in gender participation seen here has already led to changes in the approaches used by the Smiley Charity, by including girls' schools in the programme, as well as setting up separate girls' divisions within libraries. Given that girls' schools are quite numerous in Kenya, this obviously makes it possible to increase the female proportion as required.

Working in the tutor-web is primarily for learning for 74% of the students and 82% of the students want to use their SmileyCoins. Even when the student motivation is earning coins, some of those students also learn as 92% of the students claim that they learn from using it.

By far most of the students express a desire to enrol in a university. Over half of those who do not want to have not completed the relevant drillsets. A marginally higher percentage of female learners want to go to university compared to male learners.

## ACKNOWLEDGEMENTS

A large number of individuals and institutions have made this work possible. Through the years, the projects have received funding from The Icelandic Centre for Research and from several EU H2020 grants. The current initiative in the Kenyan libraries is primarily funded by the Icelandic Ministry of Foreign Affairs, including tablet purchases and corresponding logistics, but supplemented by private donations, the largest private donor being the Sliding Through Foundation.

Continuous support has been provided by the University of Iceland, including the UI Research Fund where the course material has been developed, and by the University of Iceland Science Institute where most of the research and development has been conducted. The current version of the tutor-web was developed by Jamie Lentin at Shuttle Thread Ltd and Jamie has also participated in the development of the SmileyCoin wallets. The Icelandic Research Fund has provided support for the most recent developments of the tutor-web and statistical analyses.

Countless students have contributed to the tutor-web and development of the SmileyCoin wallet.

The Smiley Charity has a subsidiary in Nairobi, where the library initiative is led by Kamau Mbugua. In addition to Kamau Mbugua, the work in the western part of Kenya is led by several individuals, particularly Zachariah Mbasu, Thomas Mawora and Maxwell Fundi.